\documentclass{article}
\usepackage[letterpaper, margin=1in]{geometry}
\usepackage{amsmath}
\usepackage{cleveref}

\DeclareRobustCommand{\rchi}{{\mathpalette\irchi\relax}}
\newcommand{\irchi}[2]{\raisebox{\depth}{$#1\chi$}} 

\newcommand\snowmass{\begin{center}\rule[-0.2in]{\hsize}{0.01in}\\\rule{\hsize}{0.01in}\\
\vskip 0.1in Submitted to the  Proceedings of the US Community Study\\ 
on the Future of Particle Physics (Snowmass 2021)\\ 
\rule{\hsize}{0.01in}\\\rule[+0.2in]{\hsize}{0.01in} \end{center}}

\begin{document}
\title{SMEFT at the LHC and Beyond: A Snowmass White Paper}
\author{William Shepherd\\Department of Physics and Astronomy, Sam Houston State University, Huntsville TX, USA\\shepherd@shsu.edu}
\date{}
\maketitle
\snowmass

\abstract{I detail an optimistic future vision for the use of SMEFT at high-energy colliders, and describe the use of the studies which could result for interpretation of future models. I also explore some of the potential pitfalls which could significantly degrade the utility of those results and discuss approaches to avoid them and ensure that results are actually accurate when applied to new models.}

\section{Introduction and Background}

In the absence of striking new physics signatures anticipated by our favorite models, the field of particle phenomenology has been pushed toward methods to determine the impact of Standard Model (SM) measurements on proposed new physics (NP) in a more model-independent fashion. The leading tool for this purpose, given its great success in multiple other regimes of physics, is Effective Field Theory (EFT). The dimensionality of operators provides a convenient ranking of the expected impact of a given effect, allowing for the use of perturbation theory in the inverse NP scale, rather than requiring detailed knowledge of the couplings and particle masses of whatever physics lies beyond the SM, to allow meaningful calculation of its impacts on the scattering of SM particles in a way that is technically feasible and systematically improvable.

There are two physically distinct and reasonable approaches to applying EFT principles to the particles of the SM, which go by the names of Higgs EFT (HEFT)\footnote{Also sometimes referred to as the Electroweak Chiral Lagrangian, or EW$\rchi$L.} and Standard Model EFT (SMEFT). The physical distinction between these two models amounts to the identity of the 125 GeV boson discovered at the LHC; the SMEFT identifies that as the Higgs boson of the SM, assumes it carries the full vacuum expectation value $v$ responsible for electroweak symmetry breaking (EWSB), and therefore assigns it to be one element of an electroweak doublet $\phi$ and enforces the full gauge symmetry of the SM on all interactions in the theory. By contrast, the HEFT does not assume that this is the sole source of EWSB. This leads to treating the theory as only being invariant under the electromagnetic gauge symmetry, and treats the would-be Goldstone bosons as completely independent of the 125 GeV scalar. This agnosticism comes with a price, however, as EWSB still must be fully explained at roughly the energy scale $v$, so there are now two power series in the theory rather than one. Both of these theories deserve detailed study, but given the striking agreement between the SM Higgs predictions and the measured properties of $h$(125) thus far, I choose here to focus on the SMEFT which identifies the two with each other.

In the SMEFT, the single new feature is an additional perturbation series in inverse powers of $\Lambda$, the characteristic mass scale of NP. This scale is a free parameter in the theory; when comparing to any specific UV theory, one should generally choose it to correspond to the mass of the lightest new particle in the UV spectrum. Because of this correspondence, fixing $\Lambda$ as an inherent feature of an analysis at the LHC is something to be avoided if at all possible; it is far better to develop bounds and likelihoods for a reasonable variety of scales. This new perturbation series has all the characteristic behaviors of any other, including those we're familiar with from e.g. gauge and/or Yukawa theories. In particular, attention should be paid to the impact of renormalization group (RG) effects on these couplings, and to the sensitivity of the numerical definition of the Lagrangian parameters in terms of physical input observables on the expansion order of the perturbation series. This dependence exists not only in the NP sector of Wilson coefficients but also feeds back into the SM coupling definitions. This last effect causes scattering amplitudes to depend even on operators which cannot appear in any graph for the process itself.

These so-called shift effects are well understood and documented~\cite{Berthier:2015oma,Brivio:2017bnu}, but the temptation to neglect them still has some adherents. They also give rise to one important source of ambiguity in interpretation of any given experimental data using the EFT, because the shift effects themselves depend on what set of data is considered to be determinative of the SM couplings, and there are multiple a priori sensible choices of data which one can make. Having made a different choice can alter the precise dependence of an observable on the Wilson coefficients of the SMEFT, and in some cases can even remove entirely the dependence on a particular operator. This complication arises, fundamentally, from our insistence on defining the would-be SM couplings prior to studying the NP effects as much as possible. A completely global view of the totality of parameter space, including e.g. the charges associated with the SM gauge group, would clarify the situation; it would then be the case that e.g. muon decay measurements were treated on par with Higgs boson production cross section, and of course the fact that both of these have effects from the SMEFT is not in any way surprising.

In this white paper, I will first lay out a vision of the function and utility of properly-performed SMEFT studies in the context of theory efforts, the LHC, and future high-energy colliders. In \cref{sec:errs} I will then address in detail the main barriers to achieving that utility, and present workable routines for addressing them on a study-by-study basis. I will then discuss how studies performed using these techniques will interplay with each other and with other measurements in \cref{sec:fits}, and conclude with a summary of my recommendations in \cref{sec:conc}. In this white paper I am focused on the use of SMEFT as a data interpretation and utility preservation tool; as such, this white paper should in no way be thought of as a complete survey of SMEFT theory work of value.

\section{The Purpose of SMEFT Studies}

There have been many different uses for the SMEFT since its popularization. These include use as an additional matching step for low energy studies of e.g. flavor changing processes, where huge amounts of data are available and multi-step running is essential to proper interpretation in the framework of UV theories, parameterizations of possible new effects in electroweak precision data (EWPD) in the famous form of the $S,T,U$ variables (though these fail to capture important possible effects on their own, only properly describing EWPD in the formally ill-defined case of so-called Universal Theories), and as motivation for and parameterization of sensitivity to generic new effects in high-energy scattering processes. Each of these purposes is well-served, but they can all be thought of as truly serving a more grand overarching vision.

The most exciting and future-oriented goal that we can address with SMEFT analyses is to have a set of measurements carried out in the SMEFT language which will enable a straightforward test of future models against LHC and other precision measurement data; this is the vision of the SMEFT as the successor to the LEP ElectroWeak Working Group analysis. Using the SMEFT for this purpose though, rather than the (pseudo-)observables themselves as was done with LEP data, introduces additional complexity.\footnote{One effort toward this observable-focused approach is the CheckMATE program~\cite{Dercks:2016npn}. This approach requires much more computing to test a single model, and doesn't scale as well with new observable introduction, but it can achieve more stringent constraints than the fully model-independent aproach using the SMEFT.} Rather than being able to calculate the observables directly in the NP model of interest and compare them to the measurements, we now proceed through the added step of matching our NP model into the SMEFT, and then comparing the SMEFT Wilson coefficients with the results of fits to the data. Nonetheless, it is essential to successfully implementing a model-independent program of this type, because otherwise the number of observables to calculate in a given model is far too unwieldy with the inclusion of all the new measurements we have made and will make. The matching can now be performed in a fully automated way (see~\cite{Proceedings:2019rnh} and citations therein), and doesn't give a significant barrier to utility, while approaching the measurements in this way avoids the need for any Monte Carlo generation to compare the NP model of interest to the plenitude of observables measured.

We can therefore truly envision a tool for comparing to precision measurements of the production, scattering, and decays of SM particles to the predictions of an arbitrary NP model which takes as input the NP Lagrangian, automatically matches it to the SMEFT, and returns the likelihood at a given parameter point in a fairly trivial way. Exploring the parameter space of such a UV model for any tension with precision measurements at lower energy would become straightforward with such a tool, and it would be an excellent guide to future model building. However, this is only truly the case if the likelihood returned by our tool accurately reflects the constraints that can be placed on the model; if the tool regularly claims overly-strict constraints it will quickly be abandoned by the community. Developing a tool which not only is easy to use but also is worth using, then, is the challenge we find ourselves facing to best make use of the SMEFT.

\section{EFT Interpretation Uncertainty in LHC Searches}
\label{sec:errs}

The true challenge that arises here is that the Wilson coefficients we measure and match to are not the only ones that can influence any given observable; in fact, there is an infinite tower of additional couplings that can also contribute to any given process, albeit at higher and higher order in $\frac{1}{\Lambda}$, and neglecting the effects of those higher-order coefficients can make our constraints much more model dependent than we initially hoped in adopting the SMEFT framework. The uncertainties inherent in this new perturbation series aren't fundamentally any different from uncertainties arising from neglect of higher-order contributions in any other perturbative calculation, but nonetheless at this time it is unfortunately common that they are neglected entirely in the analysis of scattering processes. This neglect has led to the impression that studies based in the EFT approach can always be evaded by moderately clever model building, which undermines the utility of any constraints we quote to the point that they are reasonably ignored by the model-building community.

When studies adequately address these uncertainties from missing higher order effects, though, we can make the resulting constraints far more robust to such model-building techniques. Of course, there will always be the option to build a model which includes light new particles and thus perforce evades the EFT limit entirely\footnote{Of course, this makes the model subject to direct searches for on-shell production of these new particles. It also removes the model in question from our target space explicitly, so this cannot be thought of as a failure of the EFT approach.}, but any technique which sufficiently considers the impact of those higher order effects should already be claiming no limit in the region of parameter space that such a UV model would match onto in SMEFT parameter space. Part of what enables this is having the cutoff scale of the EFT treated as a parameter in its own right, of course, and any analysis which is aware of the potential for e.g. $\frac{s}{\Lambda^2}$ contributions in the perturbation series should immediately report no bound when the `small parameter' being expanded in ceases to be less than unity.

There are two energy regimes where SMEFT studies are currently being performed, and the requirements of each of them with regard to these uncertainties are quite different. The first is the regime of low-energy, high-statistics processes, generally involving on-shell or near-threshhold production of SM particles of interest such as the Higgs boson or massive gauge bosons. In this regime, significant progress has been made in identifying all possible SMEFT contributions, not just those at leading order in perturbation theory. The other region of phase space that has been of much interest experimentally for SMEFT studies is the highest energy tails of distributions, where the statistics are significantly more suppressed but the na\"\i ve SMEFT effects are predicted to be much larger. Here I'll discuss each in turn.

\subsection{Resonant and near-threshhold processes}

Significant effort has already been invested in understanding these processes, as they are by far the most common and therefore should give us the most accurate measurements of particle properties. In particular, Higgs production and decay processes have already been studied at next to leading order in SM couplings~\cite{Corbett:2021cil}, and $Z$ boson decays have been as well~\cite{Corbett:2021eux}. The most striking feature of these observables, however, is the surprising simplicity of their dependence on SMEFT parameters.

By systematically applying equations of motion, it is possible to show that there is a finite number of different ways in which the SMEFT can contribute to any process which is dominated by three-point interactions in which all particles (including those internal to the amplitude) are on-shell. This systematically replaces derivatives in amplitudes with masses, and therefore drastically limits the number of independent kinematic contributions possible. It is possible to categorize the relevant linear combinations of Wilson coefficients that yield each NP contribution in terms of field space metrics, and as a result these results are commonly referred to as the geometric formulation of the SMEFT, or the geoSMEFT~\cite{Helset:2020yio,Hays:2020scx}. This limitation of parameters of interest leads to the opportunity to fully consider the SMEFT effects in these processes, without the need to neglect other SMEFT contributions at other orders in the EFT perturbative expansion.

This treatment effectively resums all orders of the expansion in $\frac{v}{\Lambda}$ into one coefficient for each linear combination of Wilson coefficients that can contribute to the processes in question. These resummed coefficients can be thought of as the constant term in a series expansion of a form factor function $f\left(\frac{h}{v}\right)$ which will be familiar to those who have worked in the HEFT framework, which is no accident; resumming vev contributions is in some sense equivalent to working in the HEFT formulation of a theory as well. Constraining those resummed coefficients is within reach with the experimental data we expect from the LHC. This effect is quite similar, albeit from a ground-up construction, to the proposed technique of matching in the broken phase of electroweak symmetry, also referred to as vev-improved matching~\cite{Freitas:2016iwx}. In both cases the effects of multiple Higgs vev insertions are resummed to give one effective Wilson coefficient of interest to low energy phenomena.

A similar approach is possible for amplitudes in which internal particles are off-shell or which include explicit four-point contact interactions, but it is not possible in those cases to elimiate all the higher-derivative contributions formally. Instead, the relevant kinematic invariants must be restrained to be small, of order $v$, by virtue of treating those experimental measurements as counting exercises rather than differentially in center of mass energy. This will ensure that the data is completely dominated by the threshhold region, and that indeed the relevant kinematics will remain small enough that higher derivative effects can be safely neglected. Here too, there will then be a tractable number of kinematic distributions that can be studied, allowing these data to be treated similarly to those which are truly resonance-dominated. Here too, there will be a significant resummation effect, where higher-dimensional operators can be reduced to have equivalent effects to those leading in dimensionality by sufficient vev insertion, yet the total effect can be considered as one parameter for the purpose of these studies.

The distinction between these resummed Wilson coefficients and those one might na\"\i vely analyze in the SMEFT is subtle, but potentially important. At high enough energies that the electroweak symmetry is effectively restored such a resummation is not legitimate, and in principle we should want to separate out the effects that are due to operators of different initial dimensionality\footnote{These different dimensionalities contribute differently to higher terms in the form factor series discussed above, so will give physically different predictions for scatterings that involve additional Higgs bosons or would-be Goldstones.}. However, given that our analysis is unlikely to reach beyond dimension 6 operators anytime soon, this distinction, though worthwhile to keep in mind, does not have concrete impacts on our uses of the SMEFT in the near future. 

\subsection{High-Energy tails}

The situation is entirely different at high energies, however. In these tails of the distribution we are looking at rare processes, very strongly suppressed by the parton distribution functions at large momentum fraction, with kinematic invariants much larger than $v^2$, and contributions from higher dimensional operators are both much less suppressed than in resonant phenomena and also can no longer be straightforwardly collapsed and resummed into a rescaling of the leading-operator effect. This means that considering the full EFT effect would require including all of the exponentially many operators that exist at higher operator dimensionality. In principle, we could at least utilize the extant listings of operators at dimension-8~\cite{Hays:2018zze} to more accurately calculate our signal function, but practically speaking this explodes the size of the parameter space of interest beyond any semblance of tractability without actually solving the problem of still being sensitive to the infinite tower of operators at yet-higher dimensionality. Instead, what we must do if we are interested in learning what we can at dimension 6 is to be rigorous with our treatment of the perturbation series in $\frac{s}{\Lambda^2}$.

This series is just like any other we encounter in field theory, and can be treated identically to them. The only reason this hasn't always been the standard treatment is that the analogous cases in renormalizable theories tend to demand this consistency more insistently than the current case by virtue of the presence of IR singularities. The most familiar example is QED, where the one-loop correction to the scattering vertex appears at first blush to be infinite, and that infinity is only properly regulated once additional final states are formally included in the class of scatterings being considered. Unfortunately for our understanding of EFTs, these singularities tend not to present themselves in processes where there is an additional factor of kinematic invariants multiplying the SM-like amplitude, as the relevant pole is cancelled.

The presence of the IR pole makes it perfectly manifest that one must calculate the cross section as a series in the perturbation variable, not the amplitude. If we were to do otherwise, treat the amplitude as perfectly well-known at NLO in $\alpha$, and then square that amplitude to find a cross section, the term arising from the square of the loop graph would give us an infinity which is not cancelled by the single-emission graphs that normally solve the problem of IR divergences here. This lesson applies just as well to the case of the EFT expansion. Despite the absence of infinities to provide in-your-face enforcement of this treatment of perturbation theory, it remains the correct approach.

A similar lesson which we know very well, in this case most of all from QCD processes, is the need to estimate the errors we make by truncating the perturbation series at any particular order. Generally, this is done by finding as reliable as possible an estimate of the impact of terms in the series at the subsequent order in our perturbation parameter. In the QCD case, we generally achieve this by varying the renormalization scale by some amount, as we expect that a significant portion of the next order effects will be due to corrections of the running of the coupling. There isn't a rigorous argument behind the common practice of varying the scale up and down by a factor of 2\footnote{There really couldn't be a rigorously correct choice, or the result wouldn't be an estimate.}, and we must develop a similarly somewhat ad-hoc rule for estimating our error in the EFT series truncation. Here it is again necessary to follow the usual rules of perturbation theory, so we must find an appropriate estimate of the effect of $\frac{1}{\Lambda^4}$ corrections to the cross section and then include that uncertainty in our understanding of the meaning of the calculations we have done in the SMEFT. This $\frac{1}{\Lambda^4}$ contribution to the error we are estimating here must dominate the total series truncation error whenever the EFT approximation is useful, so it is a reasonable choice as a stand-in for the full truncation uncertainty.

I have proposed and applied to dijet~\cite{Alte:2017pme,Keilmann:2019cbp} and dilepton~\cite{Alte:2018xgc,Horne:2020pot} final states at the LHC a framework to estimate these errors by using the one portion of the calculation at $\mathcal{O}\left(\frac{1}{\Lambda^4}\right)$ which we can perform with the parameters we are already studying. The terms quadratic in dimension-6 Wilson coefficients arising from the squaring of an amplitude at $\mathcal{O}\left(\frac{1}{\Lambda^2}\right)$ are well-defined; these can be thought of as analogous to the differences in a calculation at fixed order in QCD when considering different scales. These are similarly well-defined and only a contribution to the full next order result, and have long been used as an estimate of the scale of effects at the next order in perturbation theory.

In this framework, having already determined what Wilson coefficients contribute to a process, constructed orthogonal linear combinations of those parameters which correspond to experimentally distinguishable effects of the SMEFT, and selected exemplar parameters to simulate as a stand-in for any contribution of the same type, we calculate in pseudodata with identical Wilson coefficients both a signal and an error distribution, $\sigma_s,\sigma_e$, which, in {\tt MadGraph}~\cite{Alwall:2014hca} and {\tt SMEFTSim}~\cite{Brivio:2017btx} notation correspond to {\tt NP\textsuperscript{$\wedge$}2==1} and {\tt NP==1} respectively. Then, in a fit to data, we take our signal model to be defined binwise as $\sigma_{SM}+\frac{\mathcal C_{\text fit}}{\mathcal C_{\text sim}}\sigma_s\pm f\left(\mathcal C_{\text fit}\right)\sigma_e$, where $\mathcal C_{\text fit,sim}$ are the fitted and simulated values of the dimensionful Wilson coefficient, and $f\left(\mathcal C_{\text fit}\right)$ is a function chosen to rescale the error distribution to account for the dependence on multiple unknown Wilson coefficients at dimension-8. An example function would be

\begin{align}
f\left(\mathcal C_{\text fit}\right)=\frac{\mathcal C_{\text fit}^2}{\mathcal C_{\text sim}^2}+\frac{g_{SM}^2\sqrt{N_8}\sqrt{1+C_{\text fit}^2}}{\Lambda_{fit}^4\mathcal C_{\text sim}^2},
\end{align}

where $g_{SM}$ is the relevant SM coupling for the process of interest, $N_8$ is an estimate of the number of operators at dimension-8 which could contribute to such a process, and $\mathcal C_i=\frac{C_i}{\Lambda^2}$.

Note here the separation of the Wilson coefficient from the NP scale in the second term; it is important that we not allow a strong would-be bound on a dimension-6 contribution to suppress our estimate of dimensionless Wilson coefficients at higher dimensionality to be far smaller than unity - this ensures that any region of phase space where the scale of new physics is below the probed energy will be assigned interpretive errors that are large compared to the putative signal. Other functional forms which achieve similar goals may well also be reasonable estimates of our uncertainty; identifying the ideal form for this function is a potential avenue for future research.

Incorporating an estimate of the uncertainty due to missing higher order in the EFT expansion contributions to the process of interest is essential to avoiding overconfidence in our interpretations of searches at very high partonic invariant masses. Due to the higher dimensionality of the EFT operators in comparison to the SM interactions, it is generically the case that their effects grow relative to the SM with energy. In a careless analysis it is tempting to use this as a large ``lever arm'' to constrain the parameters of the SMEFT based on the absence of events at the very highest energy, but we know that this is simultaneously the region where the EFT is least reliable in its predictions.

Indeed, it has often been observed that the EFT predictions can violate unitarity at high enough energies, and frameworks have been proposed to address this problem. None of these more complicated treatments are needed, though, if we simply acknowledge the uncertainty arising from neglected terms in the EFT perturbation series, as those terms grow yet faster with energy, and will naturally reduce the weighting of these very highest energy bins in any fitting or parameter-measuring procedure.

\section{Fitting with Uncertainties}
\label{sec:fits}

The ideal fitting of any model to data is a global one, where all measurements are simultaneously considered on equal footing. It is always important that this include those measurements which we commonly think of as defining the SM itself, but that is particularly true in the SMEFT, where certain NP degrees of freedom contribute to a large number of processes solely through their corrections to the relationship between observables and would-be SM couplings and parameters. This is the approach that has long been taken, including by the LEP EWWG, but it requires a very significant investment of experimental work to appropriately combine and compare the relevant measurements between experiments and channels.

The quicker alternative is a sequential fitting, where a subset of measurements are fit first, and the output of that fit is taken as an input to a fit including more measurements. This is not ideal, but can achieve comparable results provided that the earlier fits are those using the most precise constraints, such that the subsequent measurements to be included are effectively insensitive to the variations allowed by the former fit. Given the state of measurements and calculations relevant to fitting in the SMEFT, there are actually 3 natural tiers of fits that could be performed sequentially without significant loss of fidelity; EWPD could be fit first, including the SM-defining measurements of e.g. muon lifetime, followed by measurements of on-pole properties of the Higgs and near-threshhold diboson production, and then finalized by the inclusion of high-energy tail measurements. All of the interpretation uncertainties discussed above can be modeled very well as Gaussian variables, as they correspond to the summation of contributions from multiple independent Wilson coefficients at higher dimensionality, so the Central Limit Theorem can reasonably be expected to apply.

One technical challenge that arises when fitting sequentially rather than globally is the need for care in adopting the results of the earlier fits. For example, a fit that imposes only the muon decay measurements on the theory does not actually fix any one parameter. Instead, it fixes the value of a particular combination of parameters, such that we can travel in parameter space in directions orthogonal to the constraint without violating the measurement's constraint at all. This is what gives rise to the shift effects, where operators that can contribute to one process, after defining the would-be SM parameters using input measurements, end up contributing to a slew of processes that those operators have no obvious connection to. These effects must not be neglected anytime fitting is done sequentially.

Note that the relative stringency of these observables as constraints on SMEFT parameters is dependent on error treatments; many early studies of high-energy processes have claimed very strong bounds on Wilson coefficients, but that is only due to their mistreatment of higher order effects in the SMEFT. Properly accounting for the uncertainties associated with higher dimension operators and other higher-order in perturbation theory effects translates those supposedly very strong bounds into nonexistent ones, as the dominant effect being `constrained' in those studies is actually better interpreted as an error estimate, rather than as a contribution to the well-understood signal function. The primary function of these high-energy tail measurements, properly understood, is to explore directions in parameter space to which the lower-energy, more precise measurements are insensitive, not to supercede those measurements in strength; the latter cannot be accomplished for any interestingly low-scale physics, i.e. physics potentially discoverable in the foreseeable future at new colliders, due to the inherent uncertainties of the higher order effects in the EFT perturbation series.

The only technically challenging aspect of this fitting is the explicit dependence of the error functions on the Wilson coefficients that simultaneously define the signal functions. This necessitates reevaluating those uncertainties as we travel through parameter space for our fit space, adding to computational complexity but not yielding a significant obstruction to the process. It is important that we do not fix these errors to the values predicted by the best fit point, though, as some directions in parameter space will permit fairly large excursions, and those same directions will often correspond to the ones which generate further uncertainty as you travel along them. Both of these are generic properties of parameter space directions that are constrained mainly by high-energy tails. The uncertainties must be truly treated as dynamically dependent on the Wilson coefficients and the cutoff scale of the SMEFT to adequately capture the physics of the situation and produce constraints which are accurate.

\section{Conclusions and Recommendations}
\label{sec:conc}

The time has come for working hard to encode the SM measurements we make in a way that will be legible and useful to future model builders and phenomenologists, and the SMEFT is an ideal tool for that purpose. It provides us with the needed dictionary to relate arbitrary UV physics to their impacts on low-energy behavior of SM particles, and it is capable of encoding those effects of any model with heavy new degrees of freedom. As such, investing in the infrastructure needed to best exploit the SMEFT is important to the future utility of the LHC and other high-precision particle physics data. Directions of future progress include:

\begin{itemize}

\item Continued work to explore SMEFT predictions at higher orders in would-be SM couplings for on-shell and near-threshhold observables.
\item Adoption by experiments and refinement of an error estimation scheme for high-energy observables which appropriately allows for unknown effects of higher-dimensional operators to not lead to evasion of the SMEFT bounds.
\item Development of fitting algorithms necessary to address signal-parameter-depenedent uncertainties inherent in the SMEFT structure.
\item Production of sequential fits and overall likelihood functions in parameter spaces under various flavor symmetry assumptions.
\item Moving toward establishing a fully-global fitting framework, where SM and EFT parameters are treated on equal footing.

\end{itemize}

By pushing forward along these axes, the HEP community will move toward an understanding of precision SM measurements which naturally can fold in not only precision electroweak and high-energy collider data, but also flavor and low-energy scattering probes of new physics effects in one consistent framework. These studies can yield a tool which tests any new physics model against the totality of evidence available from precision measurements and which can be trusted to constrain only those models which truly are constrained by that data.

\bibliography{Sources}

\begin{thebibliography}{10}

\bibitem{Berthier:2015oma}
L.~Berthier and M.~Trott, ``{Towards consistent Electroweak Precision Data
  constraints in the SMEFT},'' {\em JHEP}, vol.~05, p.~024, 2015.

\bibitem{Brivio:2017bnu}
I.~Brivio and M.~Trott, ``{Scheming in the SMEFT... and a reparameterization
  invariance!},'' {\em JHEP}, vol.~07, p.~148, 2017.
\newblock [Addendum: JHEP 05, 136 (2018)].

\bibitem{Dercks:2016npn}
D.~Dercks, N.~Desai, J.~S. Kim, K.~Rolbiecki, J.~Tattersall, and T.~Weber,
  ``{CheckMATE 2: From the model to the limit},'' {\em Comput. Phys. Commun.},
  vol.~221, pp.~383--418, 2017.

\bibitem{Proceedings:2019rnh}
J.~Aebischer, M.~Fael, A.~Lenz, M.~Spannowsky, and J.~Virto, eds., {\em
  {Computing Tools for the SMEFT}}, 10 2019.

\bibitem{Corbett:2021cil}
T.~Corbett, A.~Martin, and M.~Trott, ``{Consistent higher order $ \sigma
  \left(\mathcal{GG}\to h\right) $, $ \Gamma \left(h\to \mathcal{GG}\right) $
  and $\Gamma(h \to \gamma\gamma$) in geoSMEFT},'' {\em JHEP}, vol.~12, p.~147,
  2021.

\bibitem{Corbett:2021eux}
T.~Corbett, A.~Helset, A.~Martin, and M.~Trott, ``{EWPD in the SMEFT to
  dimension eight},'' {\em JHEP}, vol.~06, p.~076, 2021.

\bibitem{Helset:2020yio}
A.~Helset, A.~Martin, and M.~Trott, ``{The Geometric Standard Model Effective
  Field Theory},'' {\em JHEP}, vol.~03, p.~163, 2020.

\bibitem{Hays:2020scx}
C.~Hays, A.~Helset, A.~Martin, and M.~Trott, ``{Exact SMEFT formulation and
  expansion to $\mathcal{O}(v^4/\Lambda^4)$},'' {\em JHEP}, vol.~11, p.~087,
  2020.

\bibitem{Freitas:2016iwx}
A.~Freitas, D.~L\'opez-Val, and T.~Plehn, ``{When matching matters: Loop
  effects in Higgs effective theory},'' {\em Phys. Rev. D}, vol.~94, no.~9,
  p.~095007, 2016.

\bibitem{Hays:2018zze}
C.~Hays, A.~Martin, V.~Sanz, and J.~Setford, ``{On the impact of
  dimension-eight SMEFT operators on Higgs measurements},'' {\em JHEP},
  vol.~02, p.~123, 2019.

\bibitem{Alte:2017pme}
S.~Alte, M.~König, and W.~Shepherd, ``{Consistent Searches for SMEFT Effects
  in Non-Resonant Dijet Events},'' {\em JHEP}, vol.~01, p.~094, 2018.

\bibitem{Keilmann:2019cbp}
E.~Keilmann and W.~Shepherd, ``{Dijets at Tevatron Cannot Constrain SMEFT
  Four-Quark Operators},'' {\em JHEP}, vol.~09, p.~086, 2019.

\bibitem{Alte:2018xgc}
S.~Alte, M.~König, and W.~Shepherd, ``{Consistent Searches for SMEFT Effects
  in Non-Resonant Dilepton Events},'' {\em JHEP}, vol.~07, p.~144, 2019.

\bibitem{Horne:2020pot}
A.~Horne, J.~Pittman, M.~Snedeker, W.~Shepherd, and J.~W. Walker, ``{Shift-Type
  SMEFT Effects in Dileptons at the LHC},'' {\em JHEP}, vol.~03, p.~118, 2021.

\bibitem{Alwall:2014hca}
J.~Alwall, R.~Frederix, S.~Frixione, V.~Hirschi, F.~Maltoni, O.~Mattelaer,
  H.~S. Shao, T.~Stelzer, P.~Torrielli, and M.~Zaro, ``{The automated
  computation of tree-level and next-to-leading order differential cross
  sections, and their matching to parton shower simulations},'' {\em JHEP},
  vol.~07, p.~079, 2014.

\bibitem{Brivio:2017btx}
I.~Brivio, Y.~Jiang, and M.~Trott, ``{The SMEFTsim package, theory and
  tools},'' {\em JHEP}, vol.~12, p.~070, 2017.

\end{thebibliography}

\end{document}